\documentclass[prd,twocolumn,showpacs,amsmath,amssymb]{revtex4}
\usepackage{overpic}
\usepackage{graphicx}
\usepackage{epsfig}
\usepackage{dcolumn}
\usepackage{bm}
\usepackage{rotating}
\usepackage{multirow}
\usepackage{subfigure}
\usepackage{hhline}
\usepackage{amssymb}
\usepackage{lineno}
\usepackage{braket}
\usepackage[colorlinks,linkcolor=blue,anchorcolor=blue,citecolor=blue]{hyperref}

\begin{document}
\title{Selecting the physical solution via $\eta$-$\eta'$ mixing}

\author{Kai Zhu}
\affiliation{Institute of High Energy Physics, Beijing 100049, China}

\date{\today}

\begin{abstract}
Based on $\eta$-$\eta'$ mixing analysis, we propose a novel method to extract the physical solutions for the hadronic properties of the $Y(4230)$ resonance from the experimental data.  Experimentally, multiple solutions have been reported in the decays of $Y(4230) \to \eta J/\psi$ and $Y(4230) \to \eta' J/\psi$. Utilizing our method, we determine a unique solution for the process $Y(4230) \to \eta' J/\psi$. Likewise, two solutions for the process $Y(4230) \to \eta J/\psi$ are preferred among the originally reported three solutions under the assumption that $Y(4230)$ dose not take an $s\bar{s}$ component.
\end{abstract}

\maketitle

Interference, which corresponds to the situation in which more than one coherent amplitude is required to describe an observation, is one of the amazing features of quantum mechanics. In high energy physics experiments, the total amplitude, which is composed by individual contributions for different resonances or backgrounds, is usually adopted to describe the mass spectrum or cross sections. Sometimes, interference will affect the resonance parameters or coupling strengths substantially and then cannot be ignored. However, this feature will cause the so-called multiple-solution problem when one fits the experimental data. This means that, if one uses an amplitude square $|{\cal M}|^2=|\sum_{k} g_k e^{i\phi_k}{\cal A}_k(\theta)|^2$ fitting to data, where $g$, $\phi$, and $\theta$ are the coupling strength, relative phase, and resonance parameters, respectively, one always finds that several different sets of $(g,\phi)$ can provide an equally good fit quality ($\chi^2$ or likelihood). Therefore, there are no criteria to distinguish the $(g,\phi)$ set with physical meaning from all solutions based on a fit. This problem was recognized long ago on both the theoretical and experimental sides~\cite{0710.5627, 0707.2541, 0707.3699, Yuan:2009gd}. Furthermore, Zhu {\it et al.}~\cite{1108.2760} proved that there must be two solutions when the number of contributions are two for some specific forms, and the unknown solution can always been derived based on the known one; i.e., if one knows $(g_1, \phi_1)$, then $(g_2, \phi_2)$ can be obtained, and the reverse is also true. An analytical method to find these solutions is also provided in Ref.~\cite{1108.2760} for two contributions, and numerical methods were proposed for more than two contributions~\cite{1505.01509, 1901.01394} recently.

Mathematically, these multiple solutions are symmetric with each other; therefore, in experiments the goodness of associated fits are equal, and it is impossible to distinguish the physical solution from the others only by analyzing experimental data. Thus, ambiguity will arise when one tries to interpret the experimental results. Recently, the BESIII collaboration measured the process $e^+ e^- \to \eta' J/\psi$ and reported double solutions that are $\Gamma_{ee}B(Y(4230) \to \eta' J/\psi) = 0.06 \pm 0.03$ and $1.38\pm 0.11$ eV, respectively~\cite{1911.00885}. Even though there is a very large difference between the two solutions, the two solutions provide equally fit qualities, and it is impossible to determine which is the physical one based on a single measurement. Another example is the measurements of $e^+ e^- \to \eta J/\psi$, where the three solutions are reported as $\Gamma_{ee}B(Y(4230) \to \eta J/\psi) = 4.8 \pm 1.0$, $7.0 \pm 1.5$, and $8.0 \pm 1.7$ eV~\cite{2003.03705}. Although the divergence between these solutions is smaller than that in the $\eta' J/\psi$ channel, the ambiguity still exists and causes a large uncertainty in the interpretation of the nature of $Y(4230)$, a good tetraquark candidate with quantum numbers $I^GJ^{PC}=0^+1^{--}$~\cite{2101.10622}. It should be noted that there is a reasonable guess that there are eight solutions for the fits in Ref.~\cite{2003.03705} instead of the reported three. However, when some solutions are very close to each other, it is impossible to distinguish them experimentally. At present, we can rely only on the reported results.

In this Letter, a novel method based on an analysis of $\eta$-$\eta'$ mixing is proposed to select the physical solutions from the multiple options. The primary idea starts with a general $\eta$-$\eta'$ mixing theory in flavor SU(3), where the physical states $\eta$ and $\eta'$ are expressed as
\begin{eqnarray}
\eta &=& \mathrm{cos}\theta \eta_8 - \mathrm{sin}\theta \eta_0,  \nonumber \\
\eta' &=& \mathrm{sin}\theta \eta_8 + \mathrm{cos}\theta \eta_0,
\label{eq:mix1}
\end{eqnarray}
with mixing angle $\theta$. Here, $\eta_8$ and $\eta_0$ are the octet and singlet components of the pseudo-scalar nonet states, respectively. Then it would be instructive to readdress Eqs.~(\ref{eq:mix1}) in terms of quark components for $\eta_8$ and $\eta_0$ as~\cite{Sinha:1984qn}
\begin{eqnarray}
\eta &=& \mathrm{sin}(\theta_0 - \theta)\frac{1}{\sqrt{2}}\ket{u\bar{u}+d\bar{d}} - \mathrm{cos}(\theta_0 - \theta) \ket{s\bar{s}}, \nonumber \\
\eta' &=& \mathrm{cos}(\theta_0 - \theta)\frac{1}{\sqrt{2}}\ket{u\bar{u}+d\bar{d}} + \mathrm{sin}(\theta_0 - \theta) \ket{s\bar{s}},
\end{eqnarray}
where $\theta_0$ is the canonical mixing angle,
\begin{equation}
\theta_0 = \mathrm{arctan}\frac{1}{\sqrt{2}} \approx 35.3^\circ\;.
\end{equation}

Suppose that $Y(4230)$ only contains $c\bar{c}$ and $q\bar{q}$  quark components without $s\bar{s}$, the ratio of matrix elements between the two processes of $Y(4230) \to \eta' J/\psi$ and $Y(4230) \to \eta J/\psi$ should be written as
\begin{equation}
\left| \frac{M_{\eta'}}{M_{\eta}} \right | = \left| \frac{\mathrm{cos}(\theta_0 - \theta)}{\mathrm{sin}(\theta_0 - \theta)} \right| = \frac{\mathrm{cos}(50.1^\circ \pm 0.5^\circ)}{\mathrm{sin}(50.1^\circ \pm 0.5^\circ)} = 0.84 \pm 0.02 \;.
\end{equation}
This value is obtained by adopting the mixing angle $\theta = -14.8^\circ \pm 0.5^\circ$, which is determined by experimental measurements although an actual proper description of the $\eta$-$\eta'$ system may be given by a two-mixing angle scheme to achieve higher precision~\cite{Leutwyler:1997yr,Kaiser:1998ds}. Notice that the PDG~\cite{ParticleDataGroup:2020ssz} review claims a very uncertain value for $\theta$, i.e., $-20^\circ \sim -10^\circ$, but by averaging the results in the references quoted in this review a much more precise angle is obtained. More details of the averaging are presented in the Appendix. Substituting the ratio between the matrix elements into the ratio between the branching fractions of $B(Y(4230)\to \eta' J/\psi)$ and $B(Y(4230)\to \eta J/\psi)$, one obtains
\begin{equation}
\frac{B(Y(4230) \to \eta' J/\psi)}{B(Y(4230) \to \eta J/\psi)} = \left| \frac{M_{\eta'}}{M_{\eta}} \right |^2 \frac{\Omega_{\eta'}}{\Omega_\eta} = 0.16 \pm 0.01 \;,
\label{eq:r1}
\end{equation}
where $\Omega_{\eta'}$ and $\Omega_{\eta}$, respectively, are the phase spaces of the two processes including the P-wave effect in the decays. These phase spaces are proportional to $p^3$, in which $p$ is the momentum of $\eta$ or $\eta'$ in the $e^+ e^-$ center-of-mass frame. The ratio of the phase spaces is determined to be $\Omega_{\eta'}/\Omega_\eta=0.22$, which obviously deviates from the result that does not consider the P-wave effect. Comparing the calculated ratio of the branching fractions with experimental results, it is obvious that the solution $\Gamma_{ee}B(Y(4230) \to \eta' J/\psi) = 0.06 \pm 0.03$ eV is too small in combination with the referred solutions $\Gamma_{ee}B(Y(4230) \to \eta J/\psi) \approx 4.8$, $7.0$, and $8.0$~eV from~\cite{2003.03705}. Therefore, the other reported solution $\Gamma_{ee}B(Y(4230) \to \eta' J/\psi) = 1.38 \pm 0.11$~eV must be the physical one. Furthermore, the three different branching fraction solutions of $\Gamma_{ee}B(Y(4230) \to \eta J/\psi)$ will give ratios $0.29 \pm 0.06$, $0.20 \pm 0.05$, and $0.17 \pm 0.04$, respectively. Only the last two are consistent with the calculated one $0.16 \pm 0.01$ within $1\sigma$, so they are the preferred physical solutions.

If $Y(4230)$ also contains an $s\bar{s}$ component, Eq.~(\ref{eq:r1}) will be rewritten as
\begin{equation}
\left| \frac{M_{\eta'}}{M_{\eta}} \right |  = \left| \frac{\mathrm{cos}(\theta_0 - \theta) + \delta \mathrm{sin}(\theta_0 - \theta)}{\mathrm{sin}(\theta_0 - \theta) - \delta\mathrm{cos}(\theta_0 - \theta)} \right| \;,
\end{equation}
where $\delta$ denotes the relative weight of the $s\bar{s}$ component in $Y(4230)$ and the weight of the other components is the unit. The ratio between the branching fractions of $B(Y(4230)\to \eta' J/\psi)$ and $B(Y(4230)\to \eta J/\psi)$ will increase when $\delta$ increases. For reference, the upper limit of the ratio is estimated to be $0.47$ based on the largest possible mean value plus $3\sigma$, i.e. $0.29+3\times 0.06 = 0.47$. It amounts to $\delta = 0.4$, that can roughly be viewed as the upper limit of $\delta$.

In summary, based on an $\eta$-$\eta'$ mixing analysis, we have determined the unique physical solution of $\Gamma_{ee}B(Y(4230) \to \eta' J/\psi)$ to be $1.38 \pm 0.11$ eV. In addition, two solutions for $Y(4230) \to \eta J/\psi$ are chosen from three with an assumption that $Y(4230)$ does not take any $s\bar{s}$ component. With the present experimental results, we have also found that the $s\bar{s}$ quark component in $Y(4230)$ is limited. Improved measurements in the future will help us to determine it more precisely and to understand the nature of the charmoniumlike state $Y(4230)$ better.

\section*{ACKNOWLEDGMENTS}
 K. Z. thanks Lianjin Wu for his strong work on the measurement of $J/\psi(\psi(2S)) \to \eta' p \bar{p}$ and Zhenyu Zhang for the inspiring discussions on $\eta$-$\eta'$ mixing. This work is supported in part by National Key Research and Development Program of China under Contract No.~2020YFA0406301.

\section*{APPENDIX}
\begin{figure}[htb]
\includegraphics[width=0.45\textwidth]{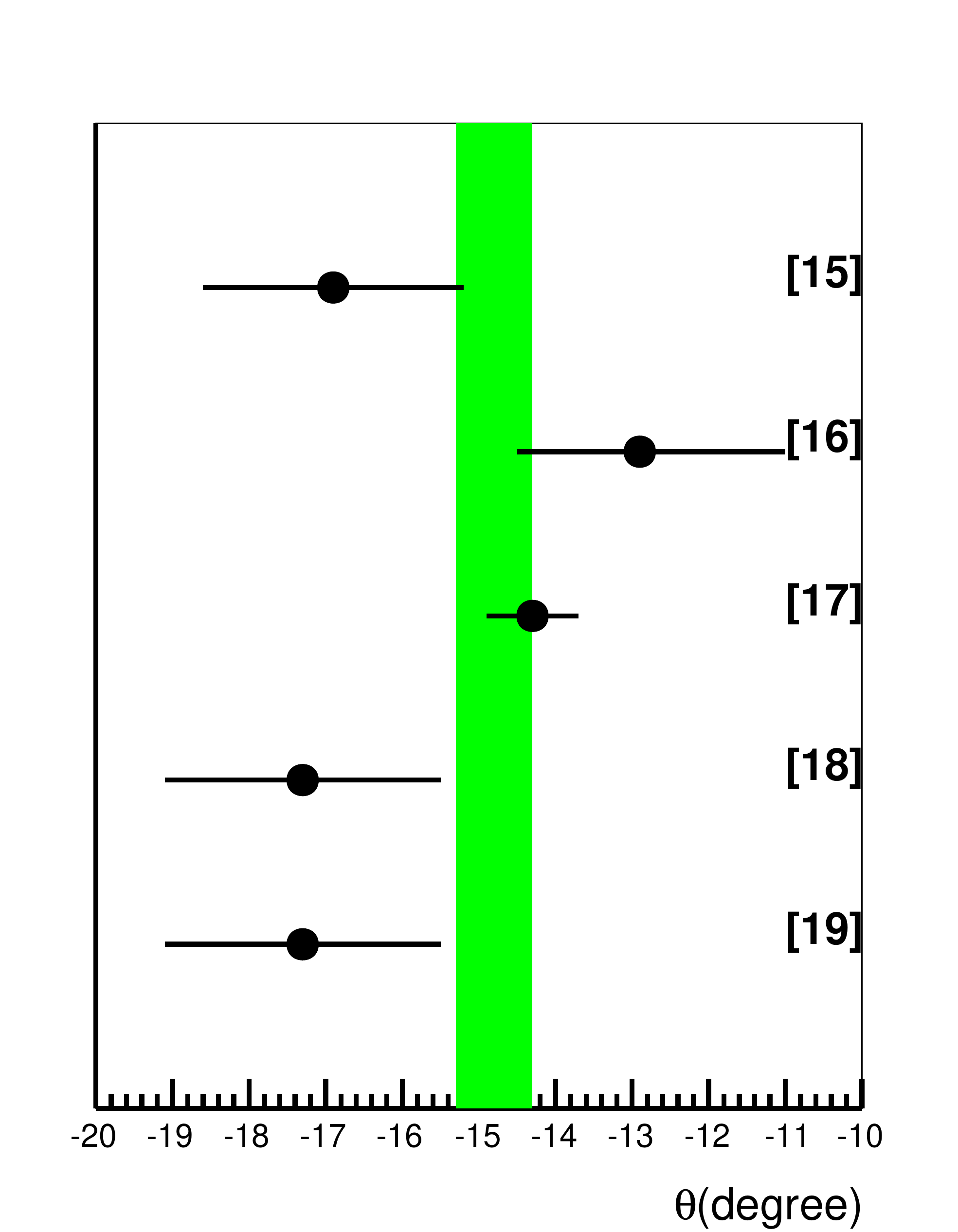}
\caption{$\eta-\eta'$ mixing angles in the references quoted by PDG~\cite{ParticleDataGroup:2020ssz} compared to the averaged one. The error bars are the results for each reference. Reference~\cite{Bramon:1997mf} is a global fit to processes $V\to \gamma P$; Ref.~\cite{KLOE:2002jed} is via $V\to\gamma P$ (not used in the average); Ref.~\cite{Ambrosino:2009sc} is via $V\to\gamma P$ and is an update of \cite{KLOE:2002jed}; Ref.~\cite{CrystalBarrel:1992ptz} is via $p\bar{p}$ annihilation; and Ref.~\cite{Amsler:1997up} is via $p\bar{p}$ annihilation, as it is a review containing the result in~\cite{CrystalBarrel:1992ptz} (which is not used in the average). The green vertical bar is the averaged mixing angle with uncertainty.}
\label{fig:ill}
\end{figure}

The PDG~\cite{ParticleDataGroup:2020ssz} review claims a very uncertain $\theta$, i.e. $-20^\circ \sim -10^\circ$. But if we fit the results in Refs.~\cite{Bramon:1997mf, KLOE:2002jed, Ambrosino:2009sc, CrystalBarrel:1992ptz, Amsler:1997up} that were quoted in~\cite{ParticleDataGroup:2020ssz} with a constant, the result is $-14.8 \pm 0.5$ with $\chi^2/NDF = 4.1/2$. This shows that the results in Refs.~\cite{Bramon:1997mf, KLOE:2002jed, Ambrosino:2009sc, CrystalBarrel:1992ptz, Amsler:1997up} are consistent with each other within $2\sigma$ and their average is much more precise than which was claimed by PDG. Related information is illustrated in Fig.~\ref{fig:ill}.

\end{document}